# Runtime Model Based Approach to Smart Home System Development

Kaidong Wu[1], Xiao He[2], Xing Chen[3], and Gang Huang[1]

[1] Peking University, Beijing 100871, China
`{wukd94, hg}@pku.edu.cn`
[2] University of Science & Technology Beijing, Beijing 100083, China
`hexiao@ustb.edu.cn`
[3] Fuzhou University, Fuzhou 350116, China
`chenxing@fzu.edu.cn`

**Abstract.** When developing smart home systems, developers integrate and compose smart devices and software applications. Because of their diversity and heterogeneity, developers usually encounter many problems. In this paper, we present a runtime model based approach to smart home system development. First, we analyze mobile applications associated with smart devices and then extract some device control APIs. Second, we use SM@RT framework to build the device runtime model. Third, we define the scenario model, that is an abstraction of devices and objects which the system consists of. Fourth, we specify mapping rules from the scenario model to the runtime model and employ a synchronizer, which can interpret the mapping rules, to keep the synchronization between the scenario model and the device runtime model. The mapping handler reads the mapping rules that are defined by developers and does the mapping in terms of them. At last, developers can program smart home systems upon the MOF-compliant scenario model using the state-of-the-art model driven technologies. https://youtu.be/SP12OtmHj50

## 1   Introduction

Smart home system development is a process of integrating and composing smart devices and various software applications (such as mobile applications and Web services [2]) into a single system to meet user requirements. In this process, developers may encounter the following problems due to the diversity and heterogeneity of smart devices:

First, developers may have to integrate and chain many smart devices, each of which has its own APIs [3]. Developers must understand the APIs of each device and know how to call them. Second, two devices that provide exactly the same function have different APIs because they are built by different manufacturers. Third, two devices that provide similar functions may have different APIs even that the two devices are built by the same manufacturer. Fourth, developers may also have to integrate some software applications and Web services whose interfaces are various.

For instance, the key requirement of an automatic planting system is to maintain a living environment for a plant. Many devices can be integrated [4], such as two smart lamps made by company A and B. Developers need to understand different APIs from different devices (and applications), and learn how to call them correctly. If developers intend to replace the smart lamp with a professional plant light that is made by company C, they have to modify the system to enable the system to control the new light.

Runtime model is a promising solution to smart home system development.

First, runtime model provides an abstraction level for devices, which enables us to handle device heterogeneity. For instance, developers can define a concept (i.e., a class) Lamp to represent both a professional plant light and a smart lamp. After that, developers can use this concept in a runtime model without caring about what concrete devices and APIs are integrated. Applications and services are seen as devices as well, so developers can use them in a consistent manner as long as they provide same functions.

Second, runtime model provides a simple way of developing smart home system. Programing at the model level based on existing model technologies (such as OCL and ATL) can ease the system development, lower the development costs and potentially reduce the time-to-market.

In this paper, we present a runtime model based approach to smart home system development. First, we analyze mobile applications associated with devices, extract APIs and use them to control devices. Second, we use SM@RT, a model-driven framework that supports model-based runtime system development, to build a runtime device model from APIs. Third, developers define a scenario model, that is an abstraction of some devices and objects which the system consists of, and set mapping rules between the scenario model and the device runtime model to synchronize them through a mapping handler. Finally, developers implement the system on the base of the scenario model. Thus, developers can pay more attention to requirements and devices, and develop the system more efficiently.

The rest of this paper is structured as follows: Section 2 shows a motivating example of the runtime model based approach to smart home system development. Section 3 introduces the framework of the approach. Section 4 provides a use case study. Section 5 discusses related work. Section 6 concludes this paper.

## 2     Motivating Example

In this section, we use an example to illustrate the problems of smart home system development.

Consider a smart planting system. Assume that we want to use some devices to automatically and intelligently take care of a plant. The system is implemented in a scenario, an environment in which developers will use the devices and objects in it to meet user requirements. We need the system to monitor the plant's living environment parameters (such as light intensity, temperature, soil moisture, soil fertility) and to determine how to maintain the plant's living environment by controlling some smart devices near the plant. For example, the system should turn on a lamp if the light intensity is too low, and turn on a water pump to water the plant when soil is very dry.

Developers will face some problems because of the heterogeneity of devices:

First, device APIs are heterogeneous. There are many devices that provide lots of functions and different APIs. Developers need to understand APIs of each device and identify the device to be used. And they can only use smart devices whose APIs are available.

Second, devices in scenarios are heterogeneous as well. For example, we use Light A from manufacturer 1 to implement the system and it runs normally. When Light B from manufacturer 2 replaces A, the system crashes, because it can't find and control Light A. Developers cannot reuse the system easily when devices or objects change.

In summary, because of the heterogeneity of devices, developers can't develop or reuse the smart home system more efficiently. If they want to redevelop a new system, they have to do what they have done in the previous system as well.

## 3   Framework

We solve the problems mentioned above by using runtime model. The overview of our approach is depicted in **Fig. 1**.

First, we get APIs from applications associated with the devices and build the device runtime mode. Other applications and services developers want to use are seen as devices as well. Developers can model the scenario and define all components they need. Then, we use mapping rules and OCL to keep the synchronization between the scenario model and the device runtime model. Finally, developers implement systems on the base of the scenario model.

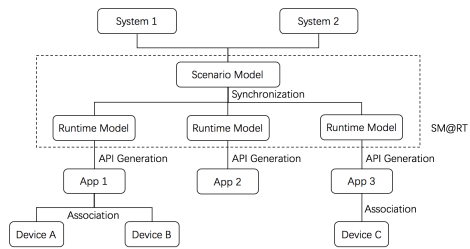

**Fig. 1.** Overview of the approach framework.

Runtime model based approach to smart home system development is shown as following steps:

### 3.1   Device API Generation

The major difference between our approach and others is that we get APIs by analyzing mobile applications rather than reading docs or introducing SDK to the project. According to previous work [1], we can extract APIs from applications, and call APIs to indirectly control devices when applications run.

For smart devices, APIs extracted from applications are customized and formatted. For non-smart devices, we can turn them on/off by controlling smart plugs associated with them. It makes building runtime model easier.

### 3.2 Runtime Model Construction

We use SM@RT framework [5] to build the runtime model on the base of device APIs.

SM@RT is a model-driven framework that supports model-based runtime system development. SM@RT can connect any systems (especially runtime system) to models to build runtime model. Other applications and services are seen as devices as well. So, we can hide the heterogeneity of devices, applications and services.

### 3.3 Scenario Model Construction

Developers need to define a scenario model, which consists of devices and objects in the scenario. The object can be abstract, like a service provider. But the objects must be able to be constructed from the data in the runtime model. For example, there isn't an air-conditioner, but we can define one, set its attribute temperature as a thermometer shows, and turn it on/off by controlling a smart electronic fan.

Scenario model can be complex, such as state machine model.

### 3.4 Mapping Rule Definition

Through a mapping handler, we can keep the synchronization between the scenario model and the device runtime model. To start it, mapping rules must be set to describe mapping relationships between models. Mapping handler will read relationships from the rules. It updates one model according to the rules when another model changes.

Mapping rules can be updated in runtime, so developers just need to modify the configuration file to let the system run again when the scenario changes.

### 3.5 System Development

Developers set rules to control the scenario model in terms of requirements, to implement the system. For a complex scenario model, developers have to set the configuration, such as transition conditions and action rules of state machine model.

## 4 Case Study

We use the development framework to develop the smart planting system above.

The major data about a plant's status consists of accumulated light (the amount of light falling on a plant in one day), temperature, soil moisture, soil fertility. We use a Flower Care (plant monitor) from MI to get the data and their suitable range of the plant, a plant light to meet the light requirement and a tiny water pump to control soil moisture. The plant light and the water pump are non-smart devices, so we use two smart plugs from MI and Haier to connect them to the system. All of the smart devices are associated with mobile applications.

The monitor has to know the plant's name, so we use an Android application named Xingse to recognize the plant by taking a picture.

The temperature indoors is usually stable within a certain range. About the soil fertility, it will be suitable if the plant is fed once per month, so we can notify user when it is too low.

We develop and run the system on a MacBook with a 2.7GHz Intel Core i5 CPU and 8 GB of RAM. We use Eclipse Neon.3 to code. Applications are installed on an Android Smart Phone with a 1.2GHz Qualcomm Snapdragon 410 CPU and 2GB of RAM.

### 4.1 Device API Generation

We extract APIs from mobile applications. These APIs are called to get the plant's data, get smart plug's status or turn it on/off and get plant's name the recognizer recognizes. They can be called when applications run.

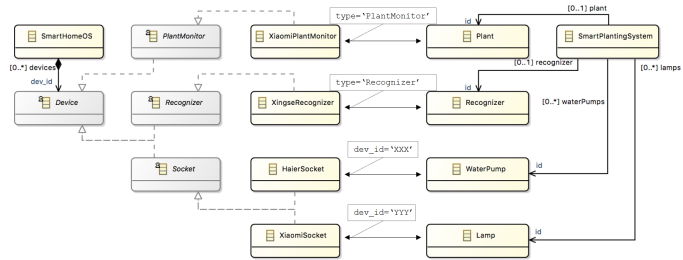

**(a)** The runtime model class diagram.   **(b)** The scenario model class diagram.

**Fig. 2.** The runtime model class diagram and the scenario model class diagram, and mapping relationships between them. All attributes and operations are hidden. These diagrams are generated by ECore Tools [6], a plugin of eclipse.

### 4.2 Runtime Model Construction

We implement the model as the standard of EMF [7], and use adapters EMF support to connect the model with the runtime system. The model's structure is shown in **Fig. 2** (a) as a class diagram.

All device runtime models are managed by *SmartHomeOS*. *Socket* represents smart plug.

### 4.3 Scenario Model Construction

The scenario model is also defined as an EMF model as shown in **Fig. 2** (b).

We suppose there are one plant and one recognizer in the scenario, and numbers of lamps and water pumps can be zero or more than 1.

In our implementation, we use only one lamp and one water pump.

### 4.4 Mapping Rule Definition

We manually set the mapping rules in a XML file, in which mapping relationships are defined like **Fig. 2**. Mapping handler will read the file and use it to keep the synchronization with OCL [8].

When updating the scenario model from the runtime model, mapping handler will update element *Plant* in the scenario model with the data in the element *Device* whose type is '*PlantMonitor*'. Because *MI Plug Mini* and *Haier UKong* are all smart plugs, we use their *dev_id*s (devices' ids), to map them respectively to a *Lamp* and a *Water-Pump* in the scenario model. Their attributes are mapped at the same time.

Device runtime models are updated with the data in associated models in the scenario model.

Developers can change the configurations dynamically to adapt the application for new requirements. So, if devices in the scenario changes, developers only need to modify the mapping rules to make the system run again.

### 4.5 System Development

We write a simple Java program. It sets the device runtime model and the scenario model, sets the mapping handler with the mapping rules and use it to manage the synchronization between models. The system opens the lamp when accumulated light is too low for the plant and opens the water pump if soil is very dry.

A web UI (see **Fig. 3**) is provided for user to set the plant's name and check the status of the plant.

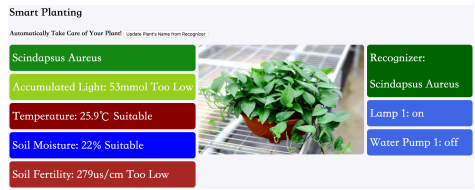

**Fig. 3.** A Screenshot of the Web UI.

Because the system is on the base of the scenario model, redevelopment and reuse of the system is easy to do.

## 5 Related Work

In smart home system development, the key problem is how to manage and use devices, sensors and other services [14, 15, 16, 17, 18, 19]. Kumar S presents a flexible standalone, low-cost smart home system based on Android applications and micro-web server [9]. Kadam R et al. introduce some technologies and tools that can be integrated in smart home systems [10]. More and more smart home systems are developed, like AirSense [11]. But in all these works, developers have to use devices providing APIs or develop components communicating with them, and focus on low-level system issues.

Runtime model can hide the heterogeneity of devices and applications, provide an abstraction level for devices. Runtime model have been used in software and IOT systems. Zhang P, et al. presents a runtime model based approach to manage intelligent campus wireless sensor [12]. API2MoL [13] is a similar approach, provides a declarative rule-based language to define mapping relationships to link API specifications and the metamodel, and automatically convert API objects into model elements or vice versa. However, developers are required to dynamically adjust systems or use the most suitable devices without APIs, which are not supported by these works.

## 6   Conclusion

We present a runtime model based approach to smart home system development. We hide the heterogeneity of device APIs through the runtime model, and hide the heterogeneity of devices in scenarios through the scenario model. The implementation of the system is based on the scenario model, so developers don't have to modify the system a lot to satisfy new requirements. The model-driven approach will make developers easier to develop smart home systems.

**References**

1. Ying Zhang, Gang Huang, Xuanzhe Liu, et.al.: Refactoring android Java code for on-demand computation offloading. In *Proceedings of Conference on Object-Oriented Programming, Systems, Languages, and Applications* (OOPSLA) Oct. 2012.
2. Xuanzhe Liu, Gang Huang, and Hong Mei. "Discovering Homogeneous Web Service Community in the User-Centric Web Environment." *IEEE Transactions on Services Computing* 2.2(2009):167-181.
3. Xuan Lu, Xuanzhe Liu, et al. "PRADA: prioritizing Android devices for apps by mining large-scale usage data." *In Proceedings of the International Conference on Software Engineering*, 2017:3-13.
4. Yun Ma, Xuanzhe Liu, et al. "Carpet: Automating Collaborative Web-Based Process across Multiple Devices by Capture-and-Replay." *Computer Software and Applications Conference* IEEE, 2015:676-685.
5. Song, H., Huang, G., Chauvel, F., Sun, Y., & Mei, H. (2010, May). SM@ RT: representing run-time system data as MOF-compliant models. In *Proceedings of the 32nd ACM/IEEE International Conference on Software Engineering-Volume 2* (pp. 303-304). ACM.
6. Ecore Tools, http://wiki.eclipse.org/Ecore_Tools.
7. Eclipse Modeling Framework (EMF), https://eclipse.org/modeling/emf/.
8. Eclipse OCL, http://wiki.eclipse.org/OCL.
9. Kumar, S. (2014).: Ubiquitous smart home system using android application. *arXiv preprint arXiv:1402.2114*.

10. Kadam, R., Mahmauni, P., & Parikh, Y. (2015).: Smart home system. *International Journal of Innovative research in Advanced Engineering (IJIRAE)*, *2*(1).
11. Fang, B., Xu, Q., Park, T., & Zhang, M. (2016, September).: AirSense: an intelligent home-based sensing system for indoor air quality analytics. In *Proceedings of the 2016 ACM International Joint Conference on Pervasive and Ubiquitous Computing* (pp. 109-119).
12. Zhang, P., & Wang, J. (2015).: Management of Intelligent Campus Wireless Sensor Networks Based on Runtime Model. *Journal of Computer and Communications*, *3*(07), 22.
13. Izquierdo, J. L. C., Jouault, F., Cabot, J., & Molina, J. G. (2012).: API2MoL: Automating the building of bridges between APIs and Model-Driven Engineering. *Information and Software Technology*, *54*(3), 257-273.
14. Liu, Xuanzhe, et al. "MUIT: A Domain-Specific Language and its Middleware for Adaptive Mobile Web-based User Interfaces in WS-BPEL." *IEEE Transactions on Services Computing* PP.99(1939):1-1.
15. Liu, Xuanzhe, et al. "Data-Driven Composition for Service-Oriented Situational Web Applications." *IEEE Transactions on Services Computing* 8.1(2015):2-16.
16. Huang G, Ma Y, Liu X, et al. Model-Based Automated Navigation and Composition of Complex Service Mashups[J]. Services Computing IEEE Transactions on, 2015, 8(1):494-506.
17. Liu Y, Liu X, Ma Y, et al. Characterizing RESTful Web Services Usage on Smartphones: A Tale of Native Apps and Web Apps[C]// IEEE International Conference on Web Services. IEEE, 2015:337-344.
18. Ma Y, Liu X, Yu M, et al. Mash Droid: An Approach to Mobile-Oriented Dynamic Services Discovery and Composition by In-App Search[C]// IEEE International Conference on Web Services. IEEE, 2015:725-730.
19. M. Brian Blake, Iman Saleh, Yi Wei, Ian D. Schlesinger, Alexander Yale-Loehr, Xuanzhe Liu. "Shared service recommendations from requirement specifications: A hybrid syntactic and semantic toolkit." Information & Software Technology 57.1(2015):392-404.